\documentclass[12pt]{article}
\usepackage{graphicx,epsfig,cite,amsmath,amssymb}
\usepackage{a4wide}
\usepackage{cite}

\newcommand{\be}{\begin{equation}}
\newcommand{\ee}{\end{equation}}
\newcommand{\ba}{\begin{eqnarray}}
\newcommand{\ea}{\end{eqnarray}}

\newcommand{\chpt}{$\chi$PT}
\newcommand{\cO}{{\cal O}}
\newcommand{\cJ}{{\cal J}}
\newcommand{\gevs}{~\mbox{GeV}^2}

\begin{document}
\begin{titlepage}
\begin{flushright}
{CAFPE-122/09}\\  {FTUV/10-0113} \\ IFIC/09-68 \\
{UG-FT-252/09} 
\end{flushright}
\vspace{2cm}
\begin{center}
{\large\bf Violation of Quark-Hadron Duality and \\[10pt] Spectral Chiral Moments in QCD}
\vfill
{\bf Mart\'{\i}n Gonz\'alez-Alonso$^a$, Antonio Pich$^a$ and Joaquim Prades$^b$}\\[0.5cm]
${}^a$
Departament de F\'{\i}sica Te\`orica and IFIC, Universitat de Val\`encia-CSIC,\\
Apt. Correus 22085, E-46071 Val\`encia, Spain\\
${}^b$ CAFPE and Departamento de F\'{\i}sica Te\'orica y del Cosmos,\\
Universidad de Granada, Campus de Fuente Nueva, E-18002 Granada, Spain\\[0.5cm]
\end{center}
\vfill
\begin{abstract}
We analyze the spectral moments of the V-A two-point correlation function.
Using all known short-distance constraints and the most recent experimental data from tau decays, we determine the
lowest spectral moments, trying to assess the uncertainties associated with the so-called violations of quark-hadron duality.  
We have generated a large number of ``acceptable'' spectral functions, satisfying all conditions, and have used them 
to extract the wanted hadronic parameters through a careful statistical analysis.
We obtain accurate values for the \chpt\ couplings $L_{10}$ and $C_{87}$, and a realistic determination
of 
the dimension six and eight contributions in the operator product expansion,
 $\cO_6= \left(-5.4\, {}^{+\, 3.6}_{-\, 1.6}\right)\cdot 10^{-3}~\mbox{GeV}^6$ and $\cO_8= \left(-8.9\, {}^{+\, 12.6}_{-\, 7.4}\right)\cdot 10^{-3}~\mbox{GeV}^8$, showing that the duality-violation effects have been underestimated in previous literature.
\end{abstract}

\vfill

\end{titlepage}

\section{Introduction}

QCD sum rules (QCDSRs) \cite{Shifman:1978bx,EdeR99} have been widely used during the last thirty years to study
many important aspects of QCD. They constitute a very useful tool,
enabling us with a powerful connection between QCD parameters and physical observables.

The basic assumption behind the QCDSR techniques is that the quark and hadron degrees of freedom
provide two dual descriptions of the same strong interaction dynamics. This quark-hadron duality is
a consequence of the assumed confinement of QCD. In more technical terms, a QCDSR is a dispersion
relation relating the value of a given two-point correlation function at some Euclidean value of $Q^2$
with an integral over the corresponding spectral function in the Minkowskian domain.
Quark-hadron duality allows us to calculate this Minkowskian integral in terms of hadrons, using the
available experimental data. Ideally, the resulting QCDSR is an exact mathematical relation
arising from analyticity and confinement (duality).
In practice, however, a series of approximations need unavoidably to be adopted in its specific
numerical implementation. In the Euclidean region the correlator is approximated by its
short-distance Operator Product Expansion (OPE) \cite{Wilson:1969zs}, truncated to a given finite
order. In the Minkowskian region, since experimental data is only available at low energies, the integral over the
physical spectral function is usually cut at a certain finite invariant-mass $s_0$; from $s_0$ up to $\infty$,
one then adopts the short-distance information provided by the OPE.

The uncertainties associated with all these approximations are usually known as violations
of quark-hadron duality. They are difficult to estimate, because of our inability to make
reliable QCD calculations at low and intermediate energies. The normal way to assess the
theoretical uncertainties of QCDSRs consists in estimating the OPE truncation error
and testing the stability of the results with variations of $s_0$. 
However, this method is too naive and can underestimate the effects not included in the OPE, i.e. the difference between the physical correlator and its OPE approximation.

Violations of QCD quark-hadron duality \cite{Shifman:2000jv} have been relatively poorly studied and often disregarded.
Its importance in finite energy sum rules (FESRs) has attracted some attention recently \cite{Cata:2005zj,GON07,Cata:2008ye,CGP09},
owing to the phenomenological need for higher accuracies. 
To estimate the size of these effects is of course of maximal importance,
if we want to master the strong interaction at all energies and be able to perform precision QCD calculations.
This importance extends to all particle physics when one realizes that those calculations are often necessary to disentangle new physics from the Standard Model. Moreover, duality violations will also be present in
new-physics scenarios characterized by a strongly-interacting dynamics. A better knowledge of duality violations
in QCD would help to understand their role in more exotic theories.

In the following, we present a detailed analysis of the possible numerical impact of duality violations
in the description of the two-point correlation function of a left- and a right-handed vector
currents. This is a very good laboratory to test the problem because this correlator is an order parameter
of chiral-symmetry breaking: in the massless quark limit it vanishes to all orders in perturbation theory;
its operator product expansion only contains power-suppressed contributions, starting with dimension six.
In the absence of any theory of duality violations, we will use
a generic, but theoretically motivated, model \cite{Cata:2008ru,Shifman:2000jv} to assess the
phenomenological relevance of these effects.

The theoretical ingredients of our analysis are presented in the next section. 
Section~\ref{sec:SpectralFunctions} contains a detailed discussion of the behaviour
of the physical spectral function at high energies. Using the most recent experimental data, we generate a 
large number of ``acceptable'' spectral functions which satisfy all known QCD constraints.
Our numerical results, obtained through a careful statistical analysis of the whole set of possible
spectral functions, are given in section~\ref{sec:numerics}. Section~\ref{sec:Summary} summarizes our findings.

\section{Theoretical Framework}

The basic objects of the theoretical analysis are the two-point correlation functions of the vector and axial-vector quark currents $\cJ_{ij}^\nu(x)$, defined as follows:
\ba
\label{eq:two}
\Pi^{\mu\nu}_{ij,\cJ}(q)
&\equiv & i \int \mathrm{d}^4 x \; \mathrm{e}^{i q x} \,
\langle 0 | T \left( \cJ_{ij}^\mu(x) \cJ_{ij}^\nu(0)^\dagger \right) | 0 \rangle
\nonumber \\ &=& (-g^{\mu\nu} q^2 + q^\mu q^\nu ) \, \Pi^{(1)}_{ij,\cJ}(q^2)
+ q^\mu q^\nu\, \Pi^{(0)}_{ij,\cJ}(q^2) \, .
\ea
Although our analysis can be applied to any correlation function, we will only study here the non-strange correlators and therefore $\cJ_{ij}^\mu(x)$ will denote the Cabibbo-allowed vector or axial-vector currents, $V_{ud}^\mu(x)=\overline{u} \gamma^\mu d$ and $A_{ud}^\mu=\overline{u} \gamma^\mu \gamma_5 d$. Moreover, we will concentrate on the $J\! =\! 0\! +\! 1$ part of the $V-A$ difference, that is nothing but 
the correlation function of the left- and right-handed currents,
$L_{ud}^\mu(x)\equiv V_{ud}^\mu(x) - A_{ud}^\mu(x)$ and $R_{ud}^\mu(x)\equiv V_{ud}^\mu(x)+A_{ud}^\mu(x)$, that is
\ba
\Pi(s)\, &\equiv&\,
\Pi_{ud,V}^{(0+1)}(s)-\Pi_{ud,A}^{(0+1)}(s)
~\equiv~
\frac{2 f_\pi^2}{s-m_\pi^2} + \overline{\Pi}(s)\, ,
\ea
where we have made explicit the contribution of the pion pole to the longitudinal axial-vector two-point function. We will work in the isospin limit, $m_u=m_d$, where $\Pi^{(0)}_{ud,V}(q^2)=0$.

\begin{figure}[tb]
\centering
\includegraphics[width=0.3\textwidth]{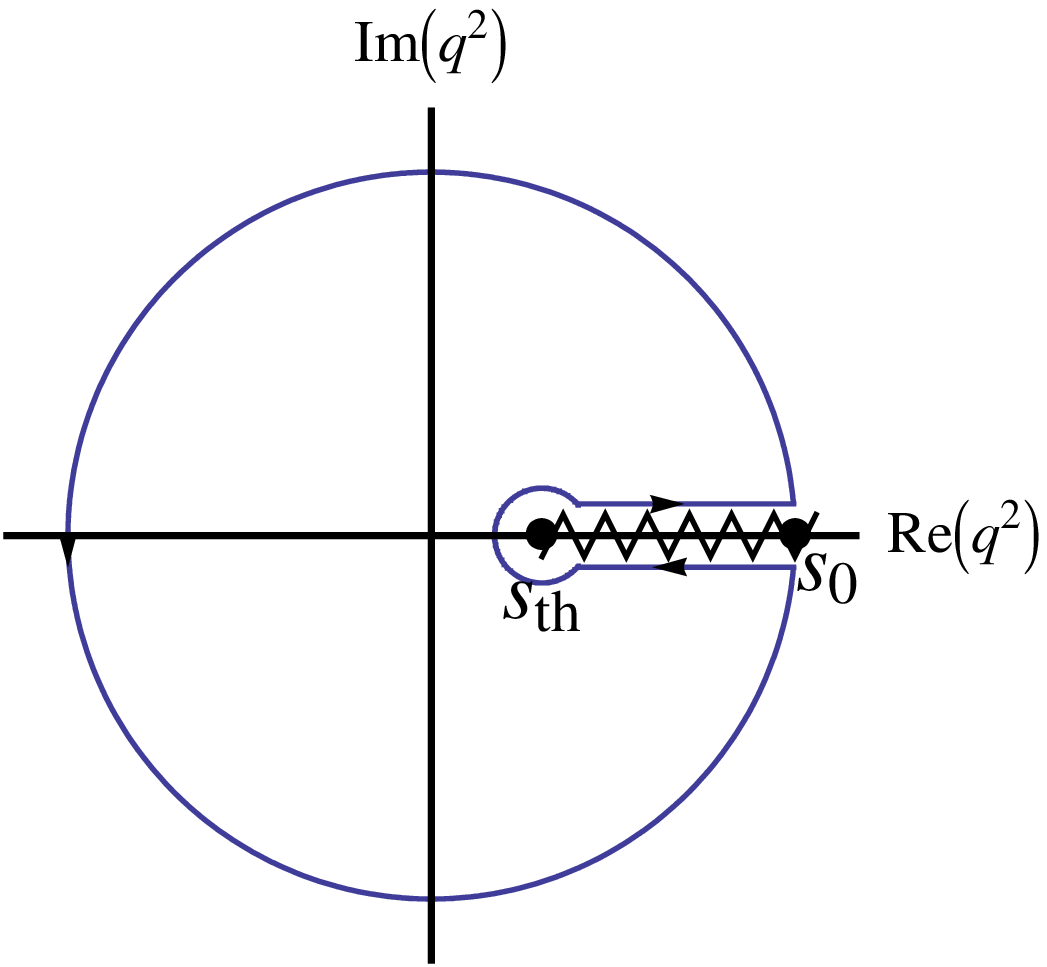}
\caption{Analytic structure of $\overline{\Pi}(s)$.}
\label{fig:circuit}
\end{figure}

The correlator $\overline{\Pi}(s)$ is analytic in the entire complex $s$-plane, except for a cut on the positive real axis which starts at the threshold $s_{\mathrm{th}}=4 m_\pi^2$. Applying Cauchy's theorem in the circuit in Fig.~\ref{fig:circuit} to the function $w(s)\,\Pi(s)$, one gets the exact relation:
\ba
\label{eq:sumrules}
\int^{s_0}_{s_{\rm th}} \mathrm{d}s\; w(s) \, \rho(s)
\, +\, \frac{1}{2 \pi i} \, \oint_{|s|=s_0} \mathrm{d}s\; w(s) \,\Pi(s)
\; =\; 2 f_\pi^2\, w(m_\pi^{2})
+ \underset{s=0}{\text{Res}} \left[ w(s) \, \Pi(s)\right] ,
\ea
where $\rho(s)\equiv \frac{1}{\pi} {\rm Im} \overline{\Pi}(s)$,
$w(s)$ is a general analytic weight function except maybe at the origin where it can have poles, and
$\underset{s=0} {\text{Res}}\, [F(s)]$ is the residue of $F(s)$ at $s=0$.
Integrals of the chiral spectral function $\rho(s)$
times $w(s)$ from threshold $s_{\rm th}$ up to $s_0$ are usually called spectral chiral moments
$M_{w(s)}(s_0)$; when $s_0 \to \infty$ we will denote them $M_{w(s)}$ for brevity.

In order to evaluate the contour integral of \eqref{eq:sumrules}, one approximates $\Pi(s)$ with
its OPE expression
\ba
\label{eq:OPE}
\Pi^{\rm OPE}(s) \; =\; \sum_{k=3}\; \frac{C_{2k}(\nu)\,\langle O_{2k}\rangle(\nu)}{(-s)^{k}}\;
 \equiv\; \sum_{k=3}\; \frac{\mathcal{O}_{2k}}{(-s)^{k}}~,
\ea
where $\langle O_{2k}\rangle(\nu)$ are vacuum expectation values
of operators with dimension $d=2k$; their associated Wilson coefficients
$C_{2k}(\nu)$ contain logarithmic dependences with $-s$. Notice that both are $\nu$-dependent quantities, but this dependence cancels in their product $\mathcal{O}_{2k}$.
The use of the OPE introduces a systematic error in the relation (\ref{eq:sumrules}), which is expected to be dominated by the region close
to the positive real axis where the OPE approximation does not apply (except at $s_0 	= \infty$ \cite{PQW76}).

Let us rewrite Eq.~\eqref{eq:sumrules} in the form:
\ba
\label{eq:SRwithDV}
\lefteqn{
\int^{s_0}_{s_{\rm th}} \!\mathrm{d}s\; w(s) \, \rho(s)
~ + ~ \frac{1}{2 \pi i} \, \oint_{|s|=s_0} \!\!\!\! \mathrm{d}s\; w(s) \,\Pi^{\rm{OPE}}(s) ~ + ~ \mathrm{DV}[w(s),s_0]}
&& \nonumber \\ && \hskip 7cm
=\; 2 f_\pi^2\, w(m_\pi^{2})
~ + ~ \underset{s=0}{\text{Res}} \left[ w(s) \, \Pi(s)\right] ,\quad
\ea
where
\ba
\label{eq:traditionalDV}
{\rm DV} [w(s),s_0]
~\equiv~ \frac{1}{2 \pi i} \,
\oint_{|s|=s_0} \mathrm{d}s\; w(s) \left( \Pi(s) - \Pi^{\rm{OPE}}(s)\right)
\ea
parameterizes the violation of quark-hadron duality that we are interested in.
Notice that ${\rm DV} [w(s),s_0]$ depends on the weight function $w(s)$ and on the circuit-radius $s_0$.
The relation~\eqref{eq:SRwithDV} contains all the elements of a standard sum rule.
The first term is the hadronic part, that in our case is nothing but an integral of the $V\!-\!A$ non-strange spectral function
that has been measured in $\tau$ decays (for $s<m_\tau^2$)
\cite{ALEPH05,Ackerstaff:1998yj,Barate:1998uf,BAR97,COA95}, while
the second term is the OPE contribution to the contour integral at $|s|=s_0$.
The second line contains the pion-pole contribution and the residue at the origin
for negative power weight functions, $1/s^n$, which is calculable with Chiral Perturbation Theory ($\chi$PT) \cite{ChPT}.

Sum rules of this type have been applied countless times in the last thirty years in order to extract theoretical parameters like quark masses \cite{mumd,ms}, the strong coupling constant \cite{alphas}, QCD condensates
\cite{Barate:1998uf,Davier:1998dz,NAR01,Bijnens:2001ps,Cirigliano:2002jy,Cirigliano:2003kc,Dominguez:2003dr,Iof06,Zyablyuk:2004iu,Rojo:2004iq,Narison:2004vz,Bordes:2005wv}  or \chpt~couplings
\cite{Davier:1998dz,KdR98,NAR01,Bordes:2005wv,GonzalezAlonso:2008rf}.
Or, used in the other way around, to make predictions of hadronic observables.

In the chiral limit ($m_u=m_d=0$) the correlator $\Pi(s)$ vanishes identically to all orders in perturbation theory and therefore its OPE contains only power-suppressed contributions from dimension $d=2k$ operators, starting at $d=6$, as we have already indicated in \eqref{eq:OPE}. The nonzero up and down quark masses induce tiny corrections with dimensions two and four, which are negligible at high values of $s$. This makes this correlator a very interesting object in the study of non-perturbative QCD.

In order to analyse duality violation (DV) effects in different sum rules, we will use the
weights $w(s)=s^n$, with $n=-2,-1,2,3$, that
generate the following four FESRs:\footnote{Here we neglect the logarithmic corrections to the Wilson coefficients in the OPE. The error associated to this approximation is expected to be smaller than the other errors involved in the analysis, as was found e.g. in Refs.~\cite{Cirigliano:2003kc,Ciulli:2003sc}.}
\ba
\label{eq:C87}
M_{-2}(s_0) \!&\equiv&\!
\int^{s_0}_{s_{\rm th}} \mathrm{d}s\; \frac{1}{s^2} \, \rho(s)\;
=\; 16 \, C_{87}^{\rm eff}\, - \, \rm{DV}[1/s^2,s_0]\, , \\
\label{eq:L10}
M_{-1}(s_0) \!&\equiv&\!
\int^{s_0}_{s_{\rm th}} \mathrm{d}s\; \frac{1}{s} \, \rho(s)\;
=\; -8 L_{10}^{\rm eff}\, - \, \rm{DV}[1/s, s_0]\, , \\
\label{eq:M2}
M_{2}(s_0) \!&\equiv&\!
\int^{s_0}_{s_{\rm th}} \mathrm{d}s\; s^2 \, \rho(s)\;
=\; 2 f_\pi^2 m_\pi^4\, +\,\cO_6 \, - \, \rm{DV}[s^2,s_0]\, , \\
\label{eq:M3}
M_{3}(s_0) \!&\equiv&\!
\int^{s_0}_{s_{\rm th}} \mathrm{d}s\; s^3 \, \rho(s)\;
=\; 2 f_\pi^2 m_\pi^6\, -\, \cO_8\, - \, \rm{DV}[s^3,s_0]\, ,
\ea
where $L_{10}^{\mathrm{eff}}\equiv -\frac{1}{8}\,\overline{\Pi}(0)$ and
$C_{87}^{\mathrm{eff}}\equiv \frac{1}{16}\,\overline{\Pi}'(0)$
are quantities that can be written in terms of low-energy \chpt~constants \cite{GonzalezAlonso:2008rf},
while $\mathcal{O}_{6,8}$ are defined in Eq.~\eqref{eq:OPE}.
These four sum rules have been used in the past \cite{Barate:1998uf,Davier:1998dz,NAR01,Bijnens:2001ps,Cirigliano:2002jy,Cirigliano:2003kc,Dominguez:2003dr,Iof06,Zyablyuk:2004iu,Rojo:2004iq,Narison:2004vz,Bordes:2005wv,KdR98,GonzalezAlonso:2008rf}
to extract the values of either the \chpt\ couplings
$L_{10}$ and $C_{87}$, or the vacuum
expectation values of the dimension six and eight operators appearing in the OPE. In those works
the DV effects  were just inferred
from the $s_0$-stability (if not just neglected),
 that as we will see can be a misleading method. Here we want to analyze
 the effect of DV on these four observables using a different approach
 that will be introduced in the following sections.

\begin{figure}[tb]
\begin{center}
\includegraphics[width=8.cm]{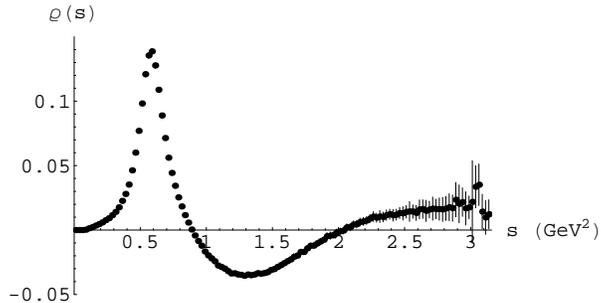}
\end{center}
\caption{Non-strange V-A spectral function $\rho(s)=\frac{1}{\pi}\mbox{Im}
\Pi^{(0+1)}_{ud,V-A}(s)$ measured from hadronic $\tau$ decays by ALEPH \cite{ALEPH05}.}
\label{fig:VmenosA}
\end{figure}

For the computation of the hadronic integral representation of the moments $M_{n}(s_0)$
we will use the 2005 ALEPH data on semileptonic $\tau$ decays \cite{ALEPH05}, shown in Fig.~\ref{fig:VmenosA},
which provide the most recent and precise measurement of the $V-A$ spectral function $\rho(s)$.

\subsection{Theoretically-known spectral moments}

In the four sum rules introduced in the previous section, we use the experimental data to extract theoretical information, namely the value of the corresponding parameters or, equivalently, the value of the spectral moments for $s_0\to\infty$, $M_n$.
There exist a few additional sum rules where we know theoretically the value of the spectral moments when $s_0\to\infty$. These sum rules will play a special role in our analysis because they give us very valuable information on the spectral function $\rho(s)$ for $s\ge s_0$.
The three sum rules that we will use are:
\ba
M_{0} &=&
\int^{\infty}_{s_{\rm th}} \mathrm{d}s\; \rho(s)
\; =\; 2 f_\pi^2\, ,
\label{eq:WSR1} \\
M_{1} &=& \int^{\infty}_{s_{\rm th}} \mathrm{d}s\; s \, \rho(s)
\; =\; 2 f_\pi^2 m_\pi^2\, ,
\label{eq:WSR2}\\
\int^{\infty}_{s_{\rm th}} \mathrm{d}s \!\!\!&s&\!\!\! \log \left({\frac{s}{\lambda^2}}\right) \,
\left. \rho(s)\right|_{m_q=0}
\; =\; (m_{\pi^0}^2-m_{\pi^+}^2)_{\mathrm{EM}} \; \frac{8\pi}{3\alpha} \, f_0^2 \, .
\label{eq:piSR}
\ea
The relations \eqref{eq:WSR1} and \eqref{eq:WSR2} are the well-known first and second Weinberg sum rules (WSRs), while the third identity is the pion sum rule ($\pi$SR) giving the electromagnetic pion mass splitting in the chiral limit \cite{Das:1967it}. In the second WSR there are contributions of the form ${\cal O}(m_q^2 \alpha_S s_0)$ \cite{Floratos:1978jb}, where $s_0$ is the upper limit of the integral, but they are negligible for the values of $s_0$ that we are considering.

\subsection{Duality violation}

To get vanishing DV in sum rules like
\eqref{eq:SRwithDV} and (\ref{eq:C87}--\ref{eq:M3})
one could think working with an infinite Cauchy radius $s_0$,
but this is clearly not an option because the spectral function $\rho(s)$
is only known up to $s_{\rm{max}}=m_\tau^2$.
We can predict the value of $\rho(s)$ at high-enough
energies using perturbative QCD, but there is an intermediate
region above $s_{\rm max}$ where perturbation theory is still not reliable.
Therefore we have to deal with this DV unavoidably, and it is important to keep in mind that at
$s_0\!\sim\! 3\gevs$ it can represent a sizable contribution to the sum rules, as the WSRs show clearly
(see e.g. Fig. 1 in Ref. \cite{Bijnens:2001ps}).

Since the solution to QCD is not known yet, DV is almost by definition a non-calculable quantity and 
that is the reason why it has been taken to be negligible very often. But
in order to make precise and reliable predictions one must worry about the size of this effect.
As it is commonly done, we have defined the DV in Eq.~\eqref{eq:traditionalDV} as the uncertainty associated with the use of the OPE.
The usual strategy to estimate the size of the DV has been to look at the stability of the spectral moments with variations of $s_0$. 
This stability can be improved adopting the so-called \lq\lq pinched weights\rq\rq
\cite{Cirigliano:2003kc}, polynomial weight functions with a zero at $s\!=\!s_0$ which suppresses the
contribution to the integral \eqref{eq:traditionalDV} from the the region close to the positive real axis.
As we will see, the stability with $s_0$ obtained with these weights can be misleading in some situations.

Taking into account that DV$[w(s),s_0]$ vanishes for $s_0\to\infty$, one can easily re-write Eq.~\eqref{eq:SRwithDV} in the form \cite{Cata:2008ru,Cata:2008ye,Cata:2005zj,GON07}
\ba
\label{eq:newDV}
\mbox{DV}	[w(s),s_0]\; =\; \int^{\infty}_{s_0} \mathrm{d}s\; w(s) \, \rho(s)~,
\ea
expressing the DV effect as an hadronic integral that can be analyzed phenomenologically.

We know from QCD that the spectral function $\rho(s)$ has to vanish at high values of $s$ and, consequently, we expect the region right above $s_0$ to be the most relevant in \eqref{eq:newDV}. This makes the ``pinched weights'' an interesting tool to minimize the DV. However, in \eqref{eq:newDV} we can see something that is hidden
in \eqref{eq:traditionalDV}, namely that one has to worry also about the possible enhancement of the contribution from the high-energy part of the integral ($s\gg s_0$) produced by the ``pinched weights''. And thus, we see that the use of these weights can worsen the situation.
Another direct consequence from \eqref{eq:newDV}, unless accidental
 cancelations occur, is that by weighting less the high-energy part
of the spectral integral one can get smaller DV.
In particular, for our spectral moments $M_{n}(s_0)$,
one expects the DV effects to increase with increasing values of $n$.
Thus, the size of the DV will be smaller in the determination
of $L_{10}^{\rm{eff}}$ than in the determination of the chiral moment
$M_2$.

To quantify the DV uncertainties of a given sum rule we must then estimate the possible
behavior of the spectral function beyond $s_0$. The DV
is an estimate of the freedom in the behavior of the spectral function above $s_0$,
once all the theoretical and phenomenological knowledge on that spectral
 function and on its moments has been taken into account.
For instance, QCD tells us that $\rho(s)$ must go \textit{quickly} enough to zero
when $s \to \infty$. This is a valuable information, but one can still imagine
infinite possible shapes for the spectral function and, therefore, the limits
imposed on DV effects are poor and not good enough for most phenomenological analyses.

Some theoretically motivated models for the DV were advocated in Ref.~\cite{Shifman:2000jv}. 
We will adopt a simple parameterization of the spectral function at high energies,
based in the resonance model proposed in \cite{Shifman:2000jv}
and similar to the one used in Refs.~\cite{Cata:2008ru,Cata:2008ye}.
Following the discussion above,
we add more physical constraints to the behaviour of $\rho(s)$ and
require that it satisfies the WSRs and the $\pi$SR \cite{GON07}.
Our goal is to generate a bunch of physically acceptable
spectral functions and translate this information into DV limits.

A similar work has been done in \cite{Cata:2008ru,Cata:2008ye}
to estimate the DV uncertainties associated with the determination of
 $\alpha_s$ from hadronic $\tau$ decay data. An important difference of our present
study with those works is that they make separate analyses for
 the vector and axial-vector channels, without imposing the constraints from the WSRs and
 $\pi$SR. In fact, one can easily check that those sum rules are not
 satisfied for the vast majority of the generated spectral functions
 used in \cite{Cata:2008ru, Cata:2008ye} (as can be seen in Fig. 2 of ref. \cite{CGP09}). So the results found there
 cannot be applied to the $V-A$ channel that we want to study here.

\section{Acceptable $V-A$ Spectral Functions}
\label{sec:SpectralFunctions}

\subsection{Spectral-function parameterization}
We split the integral of the spectral function $\rho(s)$ in two parts. For the low-energy part of the integral we will use the ALEPH data, whereas in the rest of the integration range we will work under the assumption that the spectral function is well described by the following parameterization
\ba
\label{eq:model}
\rho(s\ge s_z) &=& \kappa~ e^{-\gamma s} \sin(\beta (s-s_z))~,
\ea
that has $\kappa,\gamma,\beta$ and $s_z$ as free parameters. From the ALEPH data we know that the $V-A$ spectral function $\rho(s)$ has a second zero around $2~\mbox{GeV}^2$ (see Fig. \ref{fig:VmenosA}), which is represented in our parameterization through the $s_z$ parameter. We will take this zero as the separation point between the use of the data and the use of the model.

At high values of $s$ this parameterization appears naturally in the equidistant resonance-based model with finite widths introduced in \cite{Shifman:2000jv}. It has also been used for the vector and axial-vector
correlators in Ref.~\cite{Cata:2008ru}, based on the expected exponential fall-off associated with the intrinsic error of an asymptotic expansion; the sine function reflects the periodicity of the daughter trajectories in the spectrum of the Regge theory.

In the region $2.0~\mbox{GeV}^2\le s\le 3.3~\mbox{GeV}^2$
the proposed parameterization is compatible with the ALEPH data; the corresponding $\chi^2$ fit gives
the result\footnote{Hereafter, unless otherwise stated, we include all correlations among the points.}
\ba
\chi^2_{\mathrm{min}}(\kappa,\gamma,\beta, s_z) = \chi^2(1.00,1.05,0.40,2.03) = 4.4 ~\ll~ \mbox{d.o.f.}=43~.
\ea
In fact the compatibility appears to be too good, in the sense that the minimum $\chi^2$ is much smaller than the number of degrees of freedom (d.o.f.): 43 = 45 points - 2 parameters.
This low value of $\chi^2_{\mathrm{min}}$ was also found in Refs.~\cite{Davier:2008sk,Cata:2008ru}.

\subsection{Imposing constraints}
As we have already said, the WSRs and the $\pi$SR in \eqref{eq:WSR1}, \eqref{eq:WSR2} and \eqref{eq:piSR} are an important source of information on $\rho(s)$, for $s$ values beyond the range of the $\tau$ data. In the literature, the use of this information has been mostly limited to define the so-called \lq\lq duality points\rq\rq,
values of $s_0$ for which the WSRs are satisfied, i.e. $\mathrm{DV}[s^n,s^0]=0$ ($n=0,1$).
These duality points are frequently used to evaluate the other FESRs, but this introduces an unknown systematic error and several ambiguities, like which duality point is the best option.

We will fully use that information by imposing that the spectral function $\rho(s)$,
given by the latest ALEPH data below $s_z\!\sim\!2~\mbox{GeV}^2$ and Eq.~\eqref{eq:model} for $s\!>\!\!s_z$,
fulfils the two WSRs and the $\pi$SR within uncertainties. This requirement constrains the regions in the parameter
space of model \eqref{eq:model} that are compatible with both QCD and the data.
We will find all possible tuples\footnote{We will talk about \lq\lq tuple\rq\rq~referring
 to a set of values $(\kappa,\gamma,\beta, s_z)$.}
$(\kappa,\gamma,\beta, s_z)$ which are compatible with such constraints
 by fitting the model. In this way, we analyse
 how much freedom is left for the shape of the spectral
function after imposing all we know on $\rho(s)$ from data plus QCD.
We will also require the compatibility between models and data in the region\footnote{Although we are assuming that the model describes correctly the spectral function beyond $s_z\!\sim\!2~\mbox{GeV}^2$, we impose the compatibility with the data from $1.7~\mbox{GeV}^2$ to ensure the continuity of the spectral function in the matching region between the data and the model.} $1.7~\mbox{GeV}^2\le s\le 3.15~\mbox{GeV}^2$.

The four imposed conditions can be written quantitatively in the following form:\footnote{
The quoted errors in Eqs. \eqref{eq:1WSRcondition} and \eqref{eq:2WSRcondition} are just data errors, whereas in \eqref{eq:piSRcondition} the main uncertainty comes from the fact that quark masses do not vanish in nature and we are using real data (not chiral-limit data). We estimate this uncertainty taking for the pion decay constant the value $f_0=87\pm 5$ MeV, that covers a range that includes the physical value and the different estimates of the chiral limit value \cite{F0}. We also include a small uncertainty coming from the residual scale dependence of the logarithm, which is proportional to the second WSR. We consider $\lambda\sim 1~\mathrm{GeV}$ a good choice of scale because higher values would suppress the high-energy part of the integral (the information that we want to use), while smaller values would generate larger $\tau$-data errors in \eqref{eq:piSRcondition}, losing also information about the high-energy region.
}
\ba
\label{eq:1WSRcondition}
\lefteqn{\int_0^{s_z}\! \rho(s)^{^{\mathrm{ALEPH}}}~ds\, +\, \int_{s_z}^{\infty} \!
\rho(s;\kappa,\gamma,\beta, s_z)~ds
\; =\; 2 f_\pi^2 \; =\; \left( 17.1 \pm 0.4 \right) \cdot 10^{-3}~\mbox{GeV}^2 ,}&&
\\[10pt] \label{eq:2WSRcondition}
\lefteqn{\int_0^{s_z}\! \rho(s)^{^{\mathrm{ALEPH}}}~s~ds\, +\, \int_{s_z}^{\infty} \!
\rho(s;\kappa,\gamma,\beta, s_z)~s~ds
\; =\; 2 f_\pi^2 m_\pi^2\; =\;  \left( 0.3 \pm 0.8 \right) \cdot 10^{-3}~\mbox{GeV}^4 ,}&&
\\[10pt] \label{eq:piSRcondition}
\lefteqn{\int_0^{s_z}\! \rho(s)^{^{\mathrm{ALEPH}}}~s~\log{\left(\frac{s}{1\!\gevs}\right)}~ds\, +\,
\int_{s_z}^{\infty} \! \rho(s;\kappa,\gamma,\beta, s_z)~s~\log{\left(\frac{s}{1\!\gevs}\right)}~ds}&&
\nonumber\\
&&\hskip 3.9cm =\; (m_{\pi^0}^2-m_{\pi^+}^2)_{\mathrm{EM}} \, \frac{8\pi}{3\alpha} \, f_0^2
\; =\; -(10.9 \pm1.5) \cdot 10^{-3}~\mbox{GeV}^4  ,\qquad
\\[10pt]
\lefteqn{\chi^2(\kappa,\gamma,\beta, s_z) \; <\; \chi^2_{\mathrm{crit}}=\mbox{d.o.f.}=54\, .}&&
\ea

\subsection{Selection process of acceptable models}

After defining the minimal conditions that a tuple has to satisfy in order to be accepted, 
we perform a scanning over the 4-dimensional parameter space, looking for physically acceptable tuples.
We emphasize the importance of taking properly into account the data correlations.
For instance, if one analyses the compatibility of a null spectral function with the ALEPH data in the 
region (2, 3.15) $\gevs$, the resulting minimum $\chi^2$ is very sensitive to these correlations:
\ba
\chi^2(0.0,\gamma,\beta,s_z) / \rm{d.o.f.} &=& 0.99 ~~~~\mbox{(correlations included)}, \\
\chi^2(0.0,\gamma,\beta,s_z) / \rm{d.o.f.} &=& 4.58 ~~~~\mbox{(correlations excluded)}.
\ea

\begin{figure}[tb]
\vfill
\centerline{
\begin{minipage}[t]{.3\linewidth}\centering
\centerline{\includegraphics[width=6cm]{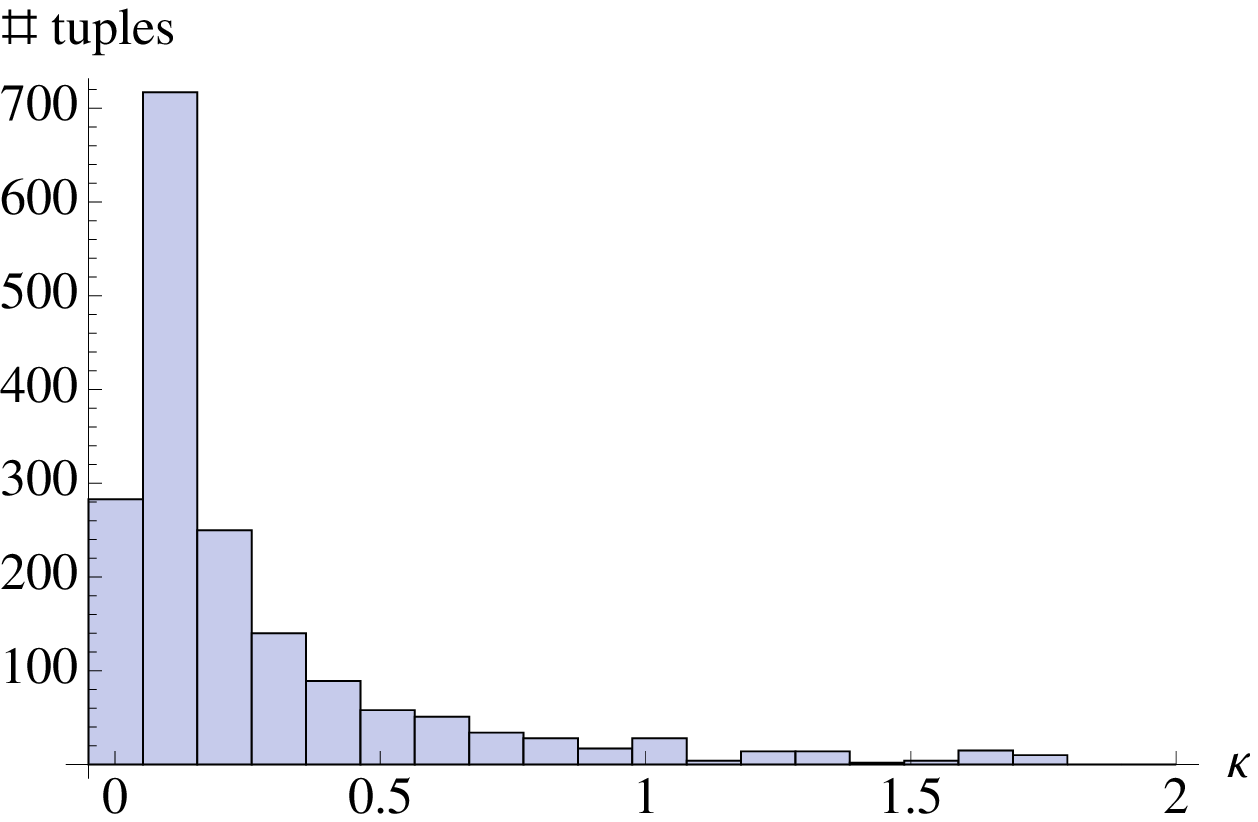}}
\end{minipage}
\hspace{2cm}
\begin{minipage}[t]{.3\linewidth}\centering
\centerline{\includegraphics[width=6cm]{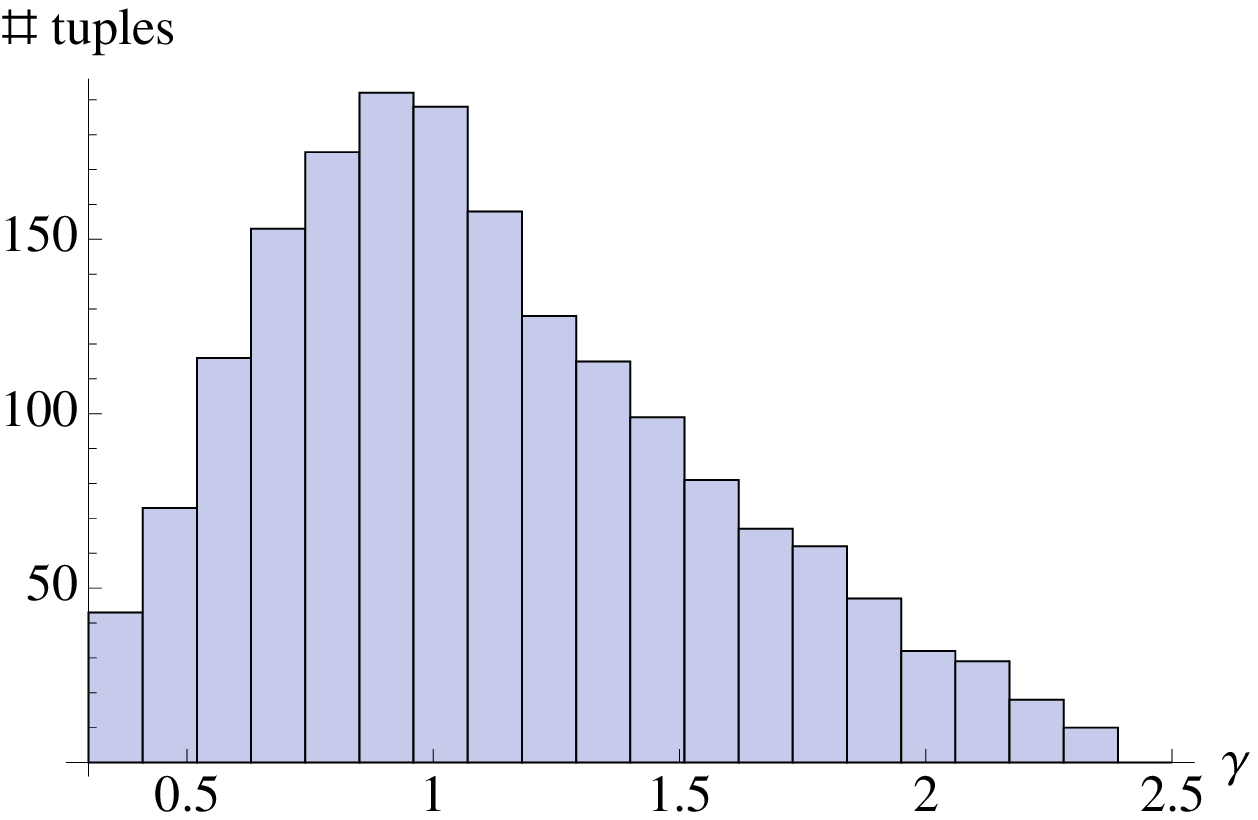}}
\end{minipage}
}
\vspace{0.7cm}
\centerline{
\begin{minipage}[t]{.3\linewidth}\centering
\centerline{\includegraphics[width=6cm]{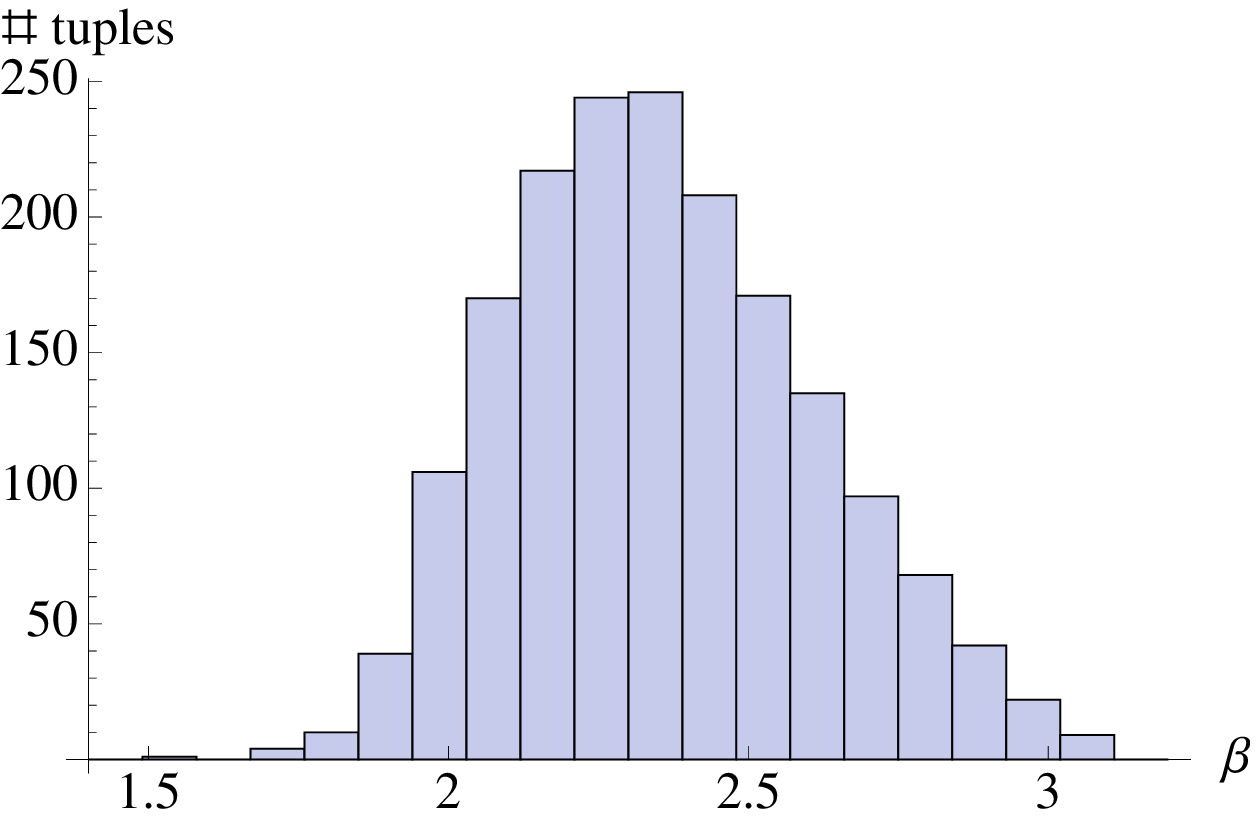}}
\end{minipage}
\hspace{2cm}
\begin{minipage}[t]{.3\linewidth}\centering
\centerline{\includegraphics[width=6cm]{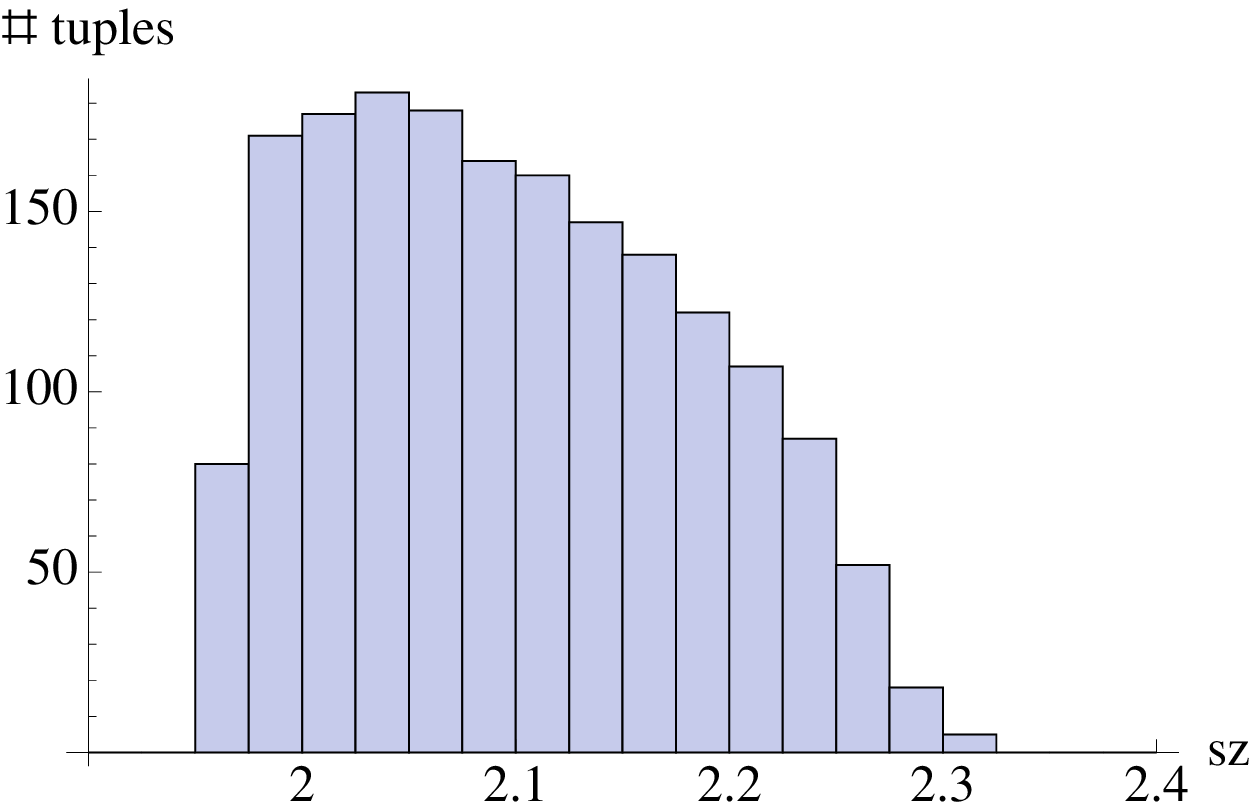}}
\end{minipage}
}
\vfill
\caption{Statistical distribution of acceptable models in the parameter space
$\kappa$ (upper-left), $\gamma$ (upper-right), $\beta$ (lower-left) and $s_z$ (lower-right).}
\label{fig:kgbshist}
\end{figure}

\begin{figure}[b]
\centering
\includegraphics[width=0.4\textwidth]{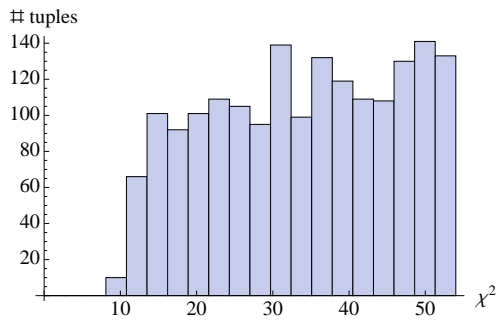}
\caption{Distribution of $\chi^2(\kappa,\gamma,\beta,s_z)$ values for acceptable tuples.}
\label{fig:chi2hist}
\end{figure}

\begin{figure}[tbh]
\vfill
\centerline{
\begin{minipage}[t]{.4\linewidth}\centering
\centerline{\includegraphics[width=7cm]{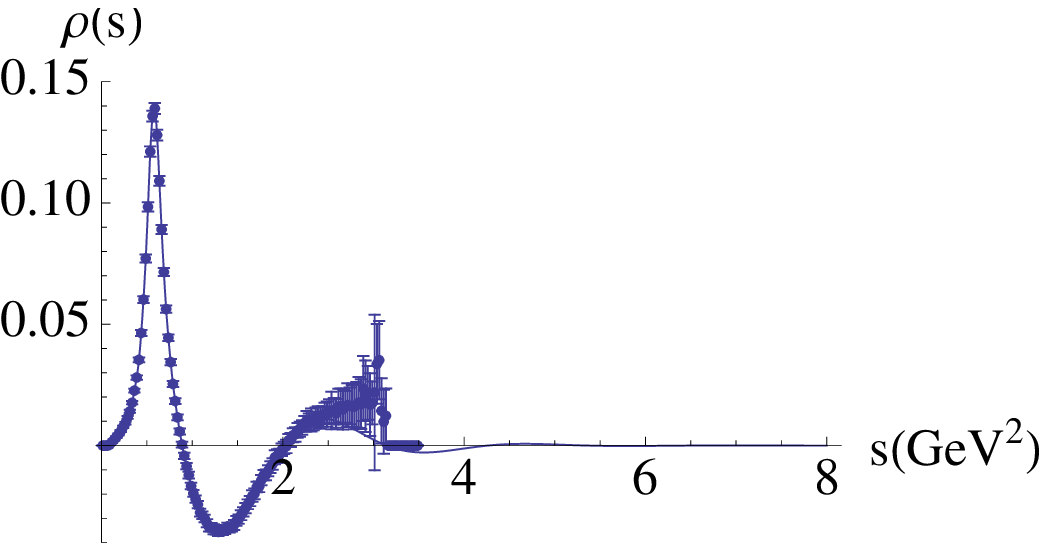}}
\end{minipage}
\hspace{2cm}
\begin{minipage}[t]{.4\linewidth}\centering
\centerline{\includegraphics[width=7cm]{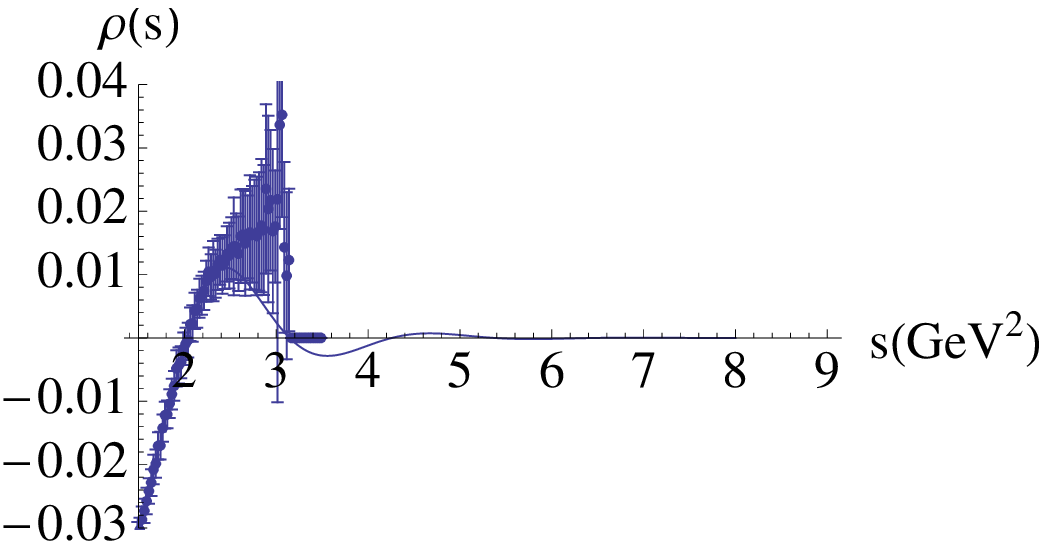}}
\end{minipage}
}
\vfill
\caption[]{Spectral function $\rho(s)$ generated with $(\kappa,\gamma,\beta,s_z)=(0.24,1.23,2.82,2.03)$, 
together with the experimental ALEPH data \cite{ALEPH05}. $\chi^2= 38.7$ for this tuple.}
\label{fig:example}
\end{figure}

To perform the
parameter-space scanning process, we adopt the following procedure.
First, we define a rectangular region such that it contains the four-dimensional ellipsoid defined by 
$\chi^2(\kappa,\gamma,\beta, s_z)=\mbox{d.o.f.}$, and we create a lattice with $20^4 = 16 \cdot 10^4$ points,
 that is, $16 \cdot 10^4$ tuples (or functions). We find that 1789 of them satisfy our set of minimal conditions; i.e., 1789 of them represent possible shapes of the physical spectral function beyond 2 $\gevs$. Fig.~\ref{fig:kgbshist}
 shows the statistical distribution of the parameters of our model after the selection process.
%
%
In Fig. \ref{fig:chi2hist} we show the distribution of the quantity $\chi^2(\kappa,\gamma,\beta,s_z)$ for those tuples that have passed the selection process. We find that all accepted tuples generate values of $\chi^2$ larger than 10.0; i.e., tuples following the central values of the experimental points do not pass the selection process; neither do the tuples that go above the central values. Thus our model indicates clearly that the third bump of the spectral function should be smaller than what the ALEPH data suggest (see Fig.~\ref{fig:VmenosA}).
The size of this third bump is an important issue that future high-quality $\tau$ decay data could clarify.
%
%
For illustrative purposes, Fig.~\ref{fig:example} shows one of the hundreds of functions that satisfy our set of conditions.

\section{Numerical Results}
\label{sec:numerics}

\begin{figure}[ht]
\vfill
\centerline{
\begin{minipage}[t]{.3\linewidth}\centering
\centerline{\includegraphics[width=6cm]{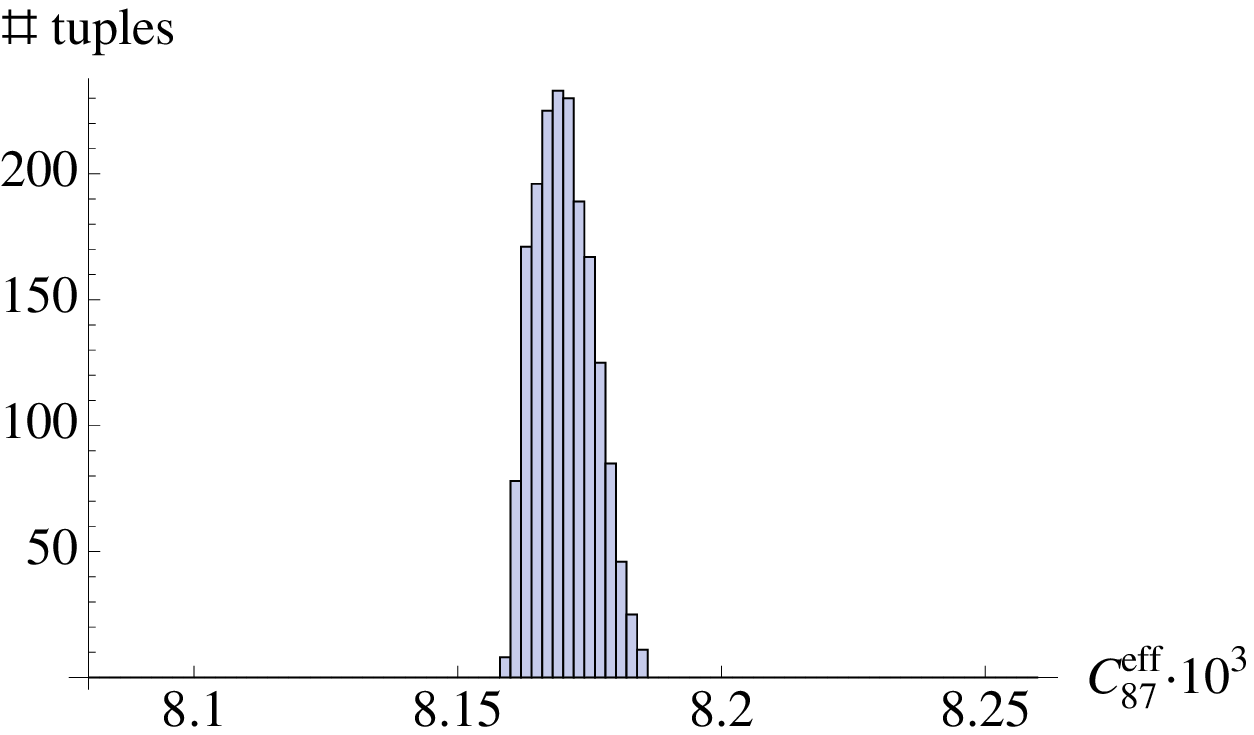}}
\end{minipage}
\hspace{2cm}
\begin{minipage}[t]{.3\linewidth}\centering
\centerline{\includegraphics[width=6cm]{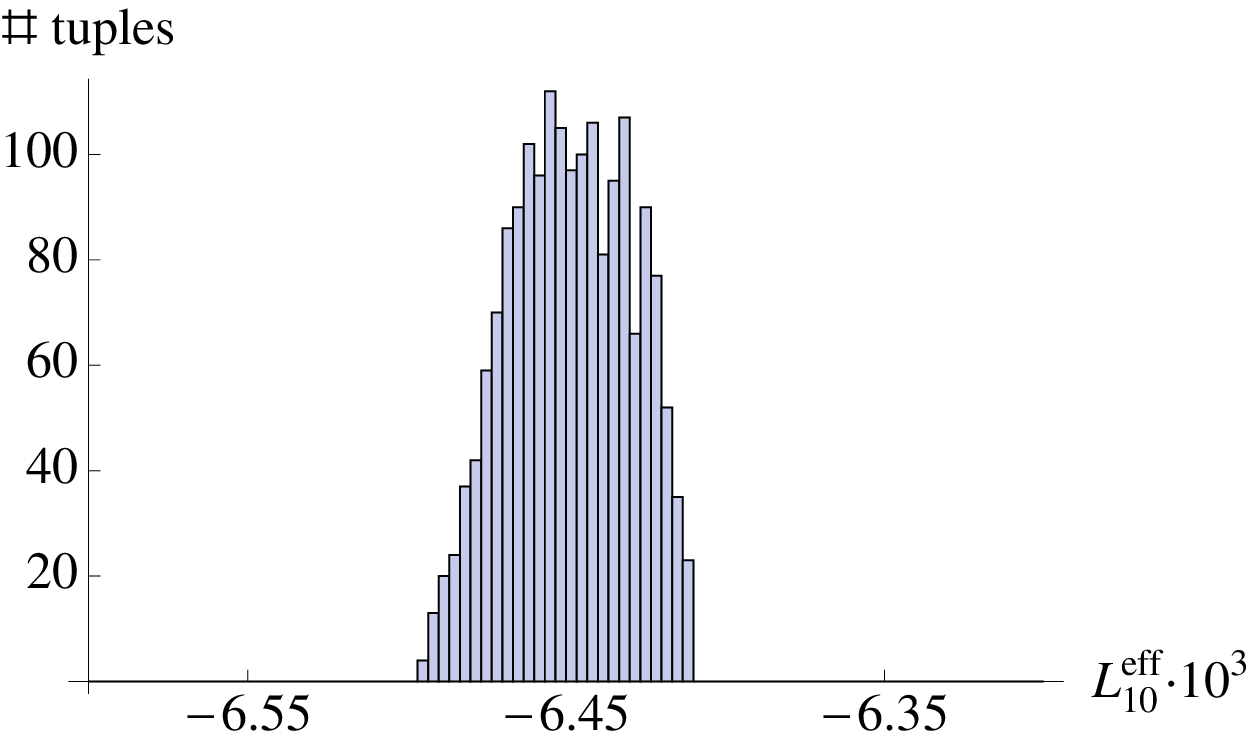}}
\end{minipage}
}
\vspace{0.7cm}
\centerline{
\begin{minipage}[t]{.3\linewidth}\centering
\centerline{\includegraphics[width=6cm]{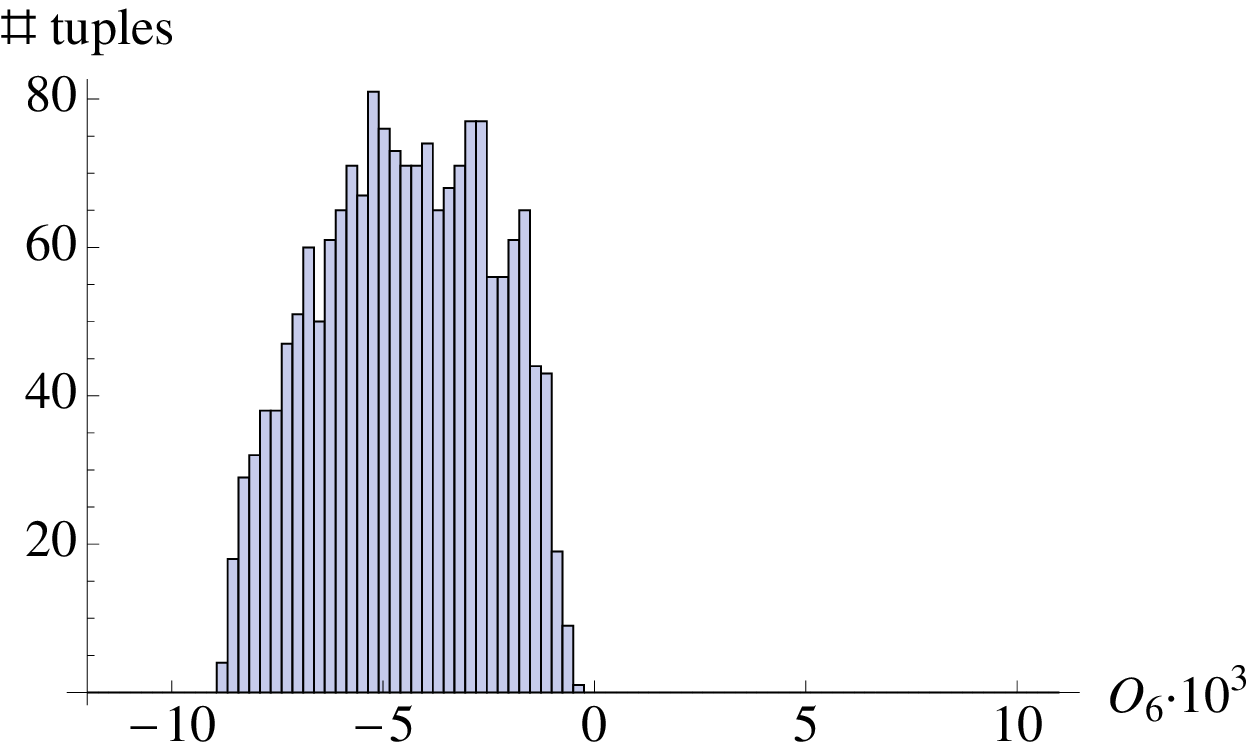}}
\end{minipage}
\hspace{2cm}
\begin{minipage}[t]{.3\linewidth}\centering
\centerline{\includegraphics[width=6cm]{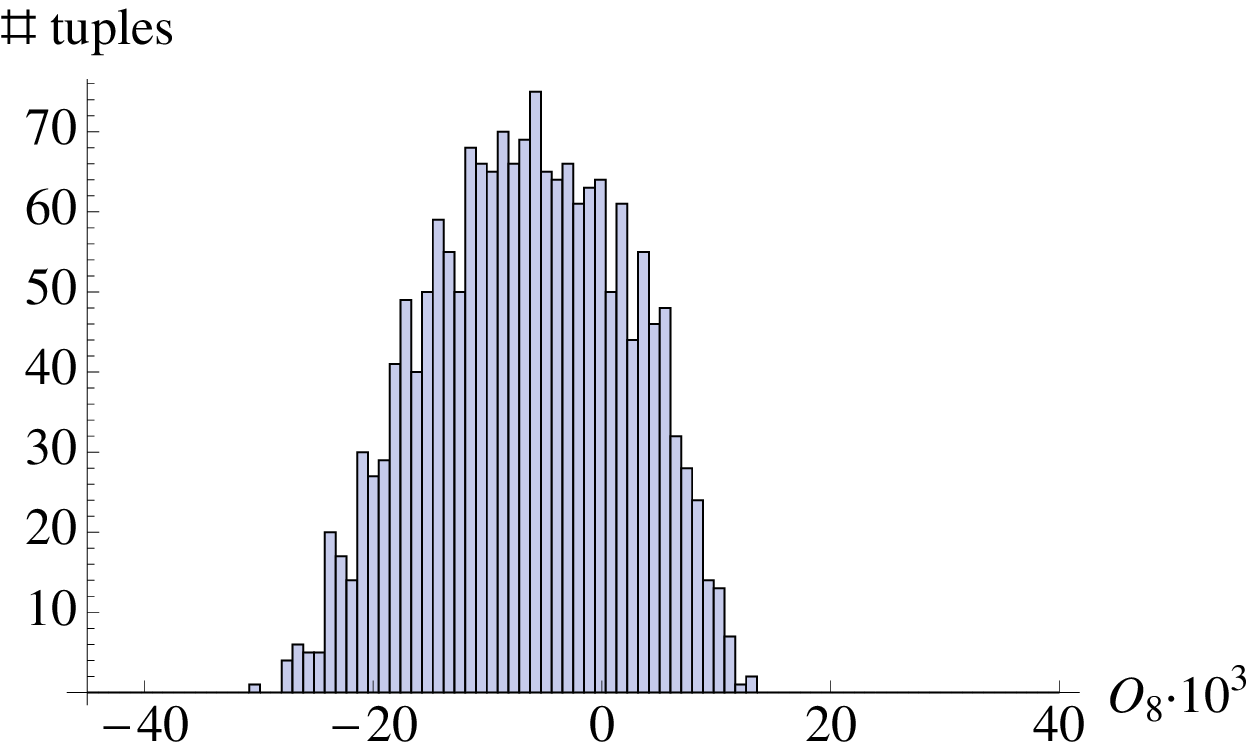}}
\end{minipage}
}
\vfill
\caption{Statistical distribution of values of
$C_{87}^{\mathrm{eff}}$ (upper-left), $L_{10}^{\mathrm{eff}}$ (upper-right), $\cO_6$ (lower-left)
and $\cO_8$ (lower-right) for acceptable models.}
\label{fig:observables}
\end{figure}

For each one of the hundreds of functions that have passed our selection process, we can calculate the associated values of $C_{87}^{\mathrm{eff}}$, $L_{10}^{\mathrm{eff}}$, $\cO_6$ and $\cO_8$, simply carrying out the integrals of Eqs.~(\ref{eq:C87}--\ref{eq:M3}) with $s_0 \to \infty$. The results of this analysis are summarized in Fig.~\ref{fig:observables}, which shows the statistical distribution of the calculated parameters. From these distributions, one gets the final numbers:
\ba
\label{eq:C87result1}
C_{87}^{\mathrm{eff}} &=& \left(8.167\, {}^{+\, 0.007}_{-\, 0.002}\pm0.12\right)\cdot 10^{-3} ~\mbox{GeV}^{-2} =
\left(8.17\,\pm0.12\right)\cdot 10^{-3}~\mbox{GeV}^{-2}~ , \\[10pt] 
L_{10}^{\mathrm{eff}}&=& \left(-6.46\, {}^{+\, 0.03}_{-\, 0.01}\pm0.07\right)\cdot 10^{-3} \;~~~~~ =~~ \;
\left(-6.46\, {}^{+\, 0.08}_{-\, 0.07}\right)\cdot 10^{-3}\, , \\[10pt] 
\cO_6&=& \left(-5.4\, {}^{+\, 3.4}_{-\, 1.0}\pm1.2\right)\cdot 10^{-3} \;~\mbox{GeV}^6 = ~\;
\left(-5.4\, {}^{+\, 3.6}_{-\, 1.6}\right)\cdot 10^{-3}~\mbox{GeV}^6\, , \\[10pt] \label{eq:O8result1}
\cO_8&=& \left(-8.9\, {}^{+\, 12.4}_{-\, 7.1}\pm2.1\right)\cdot 10^{-3} ~\mbox{GeV}^8 = ~\;
\left(-8.9\, {}^{+\, 12.6}_{-\, 7.4}\right)\cdot 10^{-3}~\mbox{GeV}^8\, ,
\ea
where the first error is that associated to the high-energy region (integral from $s_z$ to infinity), that we compute from the dispersion of the histograms of Fig. \ref{fig:observables}, and the second error is that associated to the low-energy region (integral from zero to $s_z$), that we compute in a standard way from the ALEPH data. 
This results correspond to the $68\%$ probability region (one sigma). Since the first error is not gaussian we show also now the $95\%$ probability results (95\% of the acceptable spectral functions give a result within the quoted interval):
\ba
\label{eq:C87result2}
C_{87}^{\mathrm{eff}} &=& \left(8.167\,{}^{+\, 0.011}_{-\, 0.007}\pm0.24\right)\cdot 10^{-3} ~\mbox{GeV}^{-2} =
\left(8.17\, \pm0.24\right)\cdot 10^{-3}~\mbox{GeV}^{-2}~ , \\[10pt] 
L_{10}^{\mathrm{eff}}&=& \left(-6.46\, {}^{+\, 0.04}_{-\, 0.03}\pm0.14\right)\cdot 10^{-3} \;~~~~~ =~~ \;
\left(-6.46\, {}^{+\, 0.15}_{-\, 0.14}\right)\cdot 10^{-3}\, , \\[10pt] 
\cO_6&=& \left(-5.4\, {}^{+\, 4.2}_{-\, 2.7}\pm2.4\right)\cdot 10^{-3} \;~\mbox{GeV}^6 = ~\;
\left(-5.4\, {}^{+\, 4.8}_{-\, 3.6}\right)\cdot 10^{-3}~\mbox{GeV}^6\, , \\[10pt] \label{eq:O8result2}
\cO_8&=& \left(-8.9\, {}^{+\, 16.9}_{-\, 15.1}\pm4.2\right)\cdot 10^{-3} ~\mbox{GeV}^8 = ~\;
\left(-8.9\, {}^{+\, 17.4}_{-\, 15.7}\right)\cdot 10^{-3}~\mbox{GeV}^8\, .
\ea


Our calculations have been done with a very simple, but physically motivated, parameterization of DV \cite{Cata:2008ru, Shifman:2000jv}.
Most likely this parameterization does not represent the actual shape of the $V-A$ spectral function, but
it accounts for the possible freedom of the function $\rho(s)$ beyond $2~\mbox{GeV}^2$ and its consequences on the observables.
Our statistical analysis translates the present ignorance on the high-energy behaviour of $\rho(s)$ into a clear quantitative
assessment on the uncertainties of the phenomenologically extracted parameters.

As expected, the DV effects have very little impact on the values of $C_{87}^{\mathrm{eff}}$ and $L_{10}^{\mathrm{eff}}$, because
the corresponding FESRs \eqref{eq:C87} and \eqref{eq:L10} are dominated by the low-energy region where the available data sits.
Our results are in excellent agreement with the most recent determination of these parameters, using
the same ALEPH $\tau$ data, performed in Ref.~\cite{GonzalezAlonso:2008rf}:
$C_{87}^{\mathrm{eff}} =( 8.18\pm0.14 ) \cdot 10^{-3}~\mbox{GeV}^{-2}$ \ and \
$L_{10}^{\mathrm{eff}} = - ( 6.48\pm0.06 ) \cdot 10^{-3}$.

\begin{figure}[t]
\vfill
\centerline{
\begin{minipage}[t]{.4\linewidth}\centering
\centerline{\includegraphics[width=7cm]{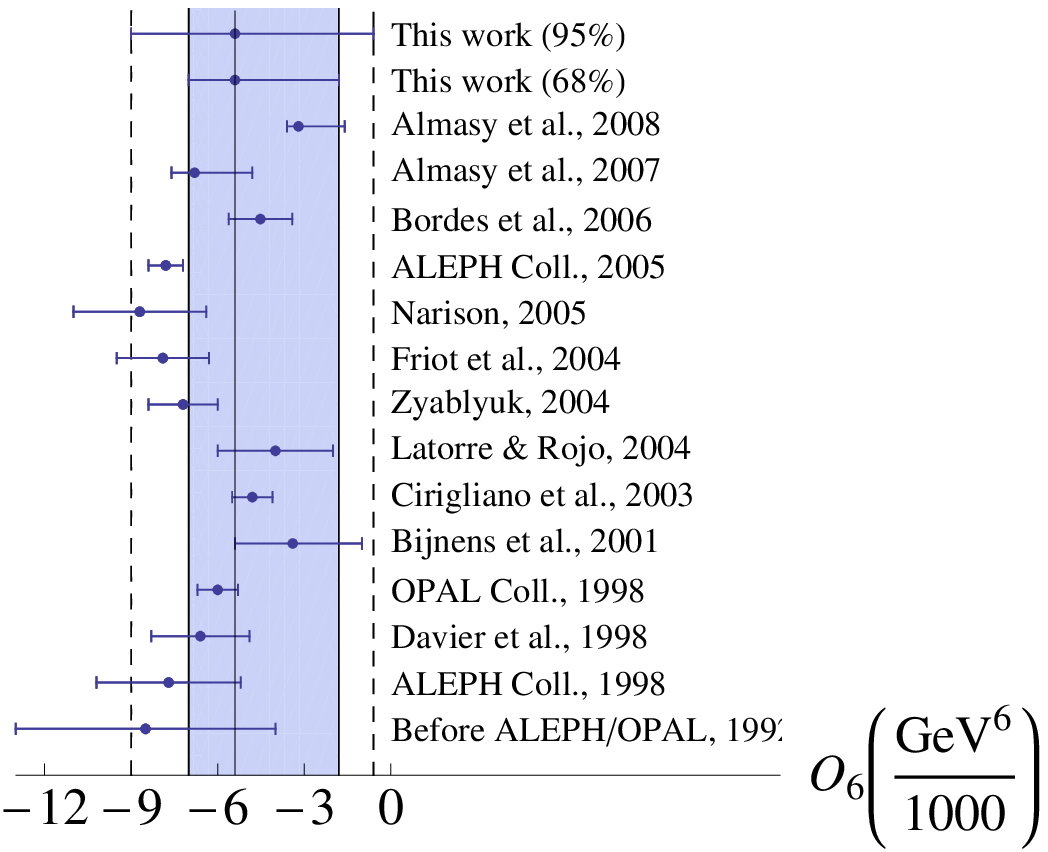}}
\end{minipage}
\hspace{2cm}
\begin{minipage}[t]{.4\linewidth}\centering
\centerline{\includegraphics[width=7cm]{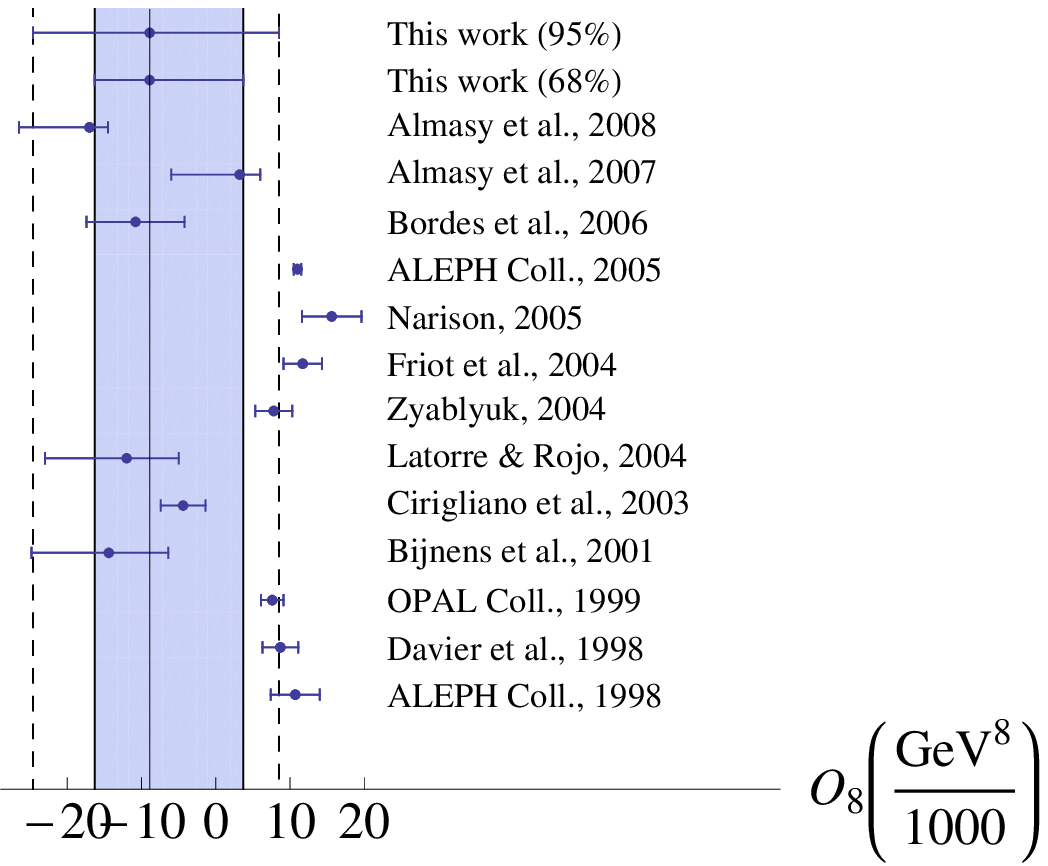}}
\end{minipage}
}
\vfill
\caption[]{Comparison of our results for $\cO_6$ (left) and $\cO_8$ (right) with previous determinations \cite{ALEPH05,Ackerstaff:1998yj,Barate:1998uf,Davier:1998dz,Bijnens:2001ps,Cirigliano:2003kc,Zyablyuk:2004iu,Rojo:2004iq,Narison:2004vz,Bordes:2005wv,deRafael,Almasy1,Almasy2} (we show for every method the most recent determination). The blue bands show our results at 65\% C.L., while the 95\% probability regions are indicated by the dotted lines. 
}
\label{fig:comparison}
\end{figure}

The situation is not so good for the moments $M_2$ and $M_3$ (or equivalently $\cO_6$ and $\cO_8$), which are sensitive to the
high-energy behaviour of the spectral function. The present ALEPH data, together with the constraints from the WSRs and the $\pi$SR,
are not good enough to determine the sign of $\cO_8$; the DV uncertainties turn out to be too large in this case.
Our results are slightly better for $\cO_6$, where there is no doubt in the sign, but again the effects of DV imply larger uncertainties
than what was estimated in previous works. Our results are compared in Fig.~\ref{fig:comparison} with previous determinations 
of $\cO_6$ and $\cO_8$.
One recognizes in the figure the existence of two groups of results that disagree between them. For $\cO_6$ there is a small tension between a bigger or smaller value, whereas in the case of $\cO_8$ the disagreement affects to the sign and is more sizeable. In some cases the discrepancy
appears to be related with a two-fold ambiguity in the adopted choice of ``duality points''.
Our analysis indicates that the DV error was grossly underestimated in most of the previous works based on FESRs \eqref{eq:C87} - \eqref{eq:M3}.
Only Refs.~\cite{Bijnens:2001ps,Rojo:2004iq} quote uncertainties similar to ours, although our error bands are slightly shifted in such a way that the tension with the other estimates is reduced.

\section{Summary}
\label{sec:Summary}

The phenomenological requirement for increasing precisions in the determinations of ha\-dronic parameters makes necessary to
assess the size of small effects which previously could be considered negligible. In particular, a substantial improvement of QCDSR results,
needed to determine many hadronic observables both in the Standard Model and in models beyond it,
could only be possible with a better control of DV.

Violations of quark-hadron duality are difficult to estimate because those effects are unknown by definition.
They originate in the uncertainties associated with the use of the OPE to approximate the exact physical correlator.
As defined in Eq.~\eqref{eq:traditionalDV}, DV effects correspond to an OPE approximation performed in the complex plane,
outside the Minkowskian region, which deteriorates in the vicinity of the real axis. Using analyticity, the size of DV can be related
with an integral of the hadronic spectral function from $s_0$ up to $\infty$, given in Eq.~\eqref{eq:newDV}, which allows us
to perform a phenomenological analysis.

We have studied the possible role of DV in the two-point correlation function $\Pi(s)$. This $V-A$ non-strange correlator is very
well suited for this analysis because: i) it is a purely non-perturbative quantity in the chiral limit,
ii) there are well-known theoretical constraints, and iii) there exist good available data from $\tau$ decays.
Moreover, different moments of its spectral function provide hadronic parameters of high phenomenological relevance.

We have assumed a generic, but theoretically motivated, behaviour of the spectral function at high energies, where data are not available,
with four free parameters. This allows us to study how much freedom in $\rho(s)$ could be tolerated, beyond the requirement that all known
QCD constraints are satisfied. Performing a numerical scanning over the four-dimensional parameter space, we have generated a large number
of ``acceptable'' spectral functions, satisfying all conditions, and have used them to extract the wanted hadronic parameters through a
careful statistical analysis. The dispersion of the numerical results provides then a good quantitative assessment of the actual
uncertainties.

We have determined four hadronic parameters of special interest:
$C_{87}^{\mathrm{eff}}$, $L_{10}^{\mathrm{eff}}$, $\cO_6$ and $\cO_8$. Our final numerical results are given in Eqs.~\eqref{eq:C87result1}-\eqref{eq:O8result1}
 for the one sigma  results and \eqref{eq:C87result2}-\eqref{eq:O8result2}
for  the 95 \% probability results. 
The  parameters $C^{\rm eff}_{87}$ and $L^{\rm eff}_{10}$
 are in excellent agreement with
 the most recent determination using FESRs and the same ALEPH data 
\cite{GonzalezAlonso:2008rf}. 
The vacuum condensate $\cO_6$ is an  important input for
the calculation of the CP-violating kaon parameter $\varepsilon'_K$,
 it dominates the $\Delta I=3/2$ contribution to $\varepsilon'_K$
\cite{Bijnens:2001ps,Cirigliano:2002jy}.
 The determination of this contribution is an important goal of lattice QCD calculations and independent information is required to
test the reliability of those results. We will study the consequences of our results for $\varepsilon'_K$ in a forthcoming publication \cite{GPP2009}.

Our analysis indicates that the DV error was grossly underestimated in most of the previous QCDSR determinations of $\cO_6$ and $\cO_8$ based on the FESRs \eqref{eq:C87} - \eqref{eq:M3}.
The present $V-A$ non-strange tau data between 2 GeV$^2$ and 3 GeV$^2$ \cite{ALEPH05} is not good enough to constrain the
spectral function with the needed accuracy. Good data in that energy region with much smaller experimental uncertainties is clearly required.
Future high-statistics $\tau$-decay data samples could allow a substantial improvement of our results, helping to clarify the actual high-energy
behaviour of the $V-A$ spectral function.

\section*{Acknowledgments}
This work has been supported in part by the EU MRTN network FLAVIAnet [Contract No. MRTN-CT-2006-035482], by MICINN, Spain 
[Grants FPA2007-60323 (M.G.-A., A.P), FPA2006-05294 (J.P.) and
Consolider-Ingenio 2010 Program CSD2007-00042 --CPAN--] and by Junta de Andaluc\'{\i}a (J.P.) [Grants P07-FQM 03048 and P08-FQM 101].
The work of M.G.-A. is funded through an FPU Grant (MICINN, Spain).

\end{document}